\documentclass[%
 reprint,
superscriptaddress,
 amsmath,amssymb,
 aps,
 prr,
]{revtex4-1}

\usepackage{color}
\usepackage[normalem]{ulem}
\usepackage{placeins}
\usepackage{verbatim}
\usepackage{graphicx}% Include figure files
\usepackage{dcolumn}% Align table columns on decimal point
\usepackage{bm}% bold math
\begin{document}

\preprint{APS/123-QED}

\title{Neural Network Solutions to Differential Equations in Non-Convex Domains: \\Solving the Electric Field in the Slit-Well Microfluidic Device}% Force line breaks with \\

\author{Martin Magill}
\affiliation{%
 Faculty of Science,
 University of Ontario Institute of Technology,
 2000 Simcoe St N, 
 Oshawa, Ontario, Canada L1H7K4
 }
\affiliation{%
 Vector Institute,
 661 University Ave Suite 710,
 Toronto, Ontario, Canada M5G1M1
 }
\author{Andrew M. Nagel}
\affiliation{%
 Faculty of Science,
 University of Ontario Institute of Technology,
 2000 Simcoe St N, 
 Oshawa, Ontario, Canada L1H7K4
 }
\author{Hendrick W. de Haan}%
 \email{Hendrick.deHaan@uoit.ca}
\affiliation{%
 Faculty of Science,
 University of Ontario Institute of Technology,
 2000 Simcoe St N, 
 Oshawa, Ontario, Canada L1H7K4
 }%

\date{\today}% It is always \today, today,

\widetext
\begin{abstract}
The neural network method of solving differential equations is used to approximate the electric potential and corresponding electric field in the slit-well microfluidic device.
The device's geometry is non-convex, making this a challenging problem to solve using the neural network method.
To validate the method, the neural network solutions are compared to a reference solution obtained using the finite element method.
Additional metrics are presented that measure how well the neural networks recover important physical invariants that are not explicitly enforced during training: spatial symmetries and conservation of electric flux.
Finally, as an application-specific test of validity, neural network electric fields are incorporated into particle simulations.
Conveniently, the same loss functional used to train the neural networks also seems to provide a reliable estimator of the networks' true errors, as measured by any of the metrics considered here.
In all metrics, deep neural networks significantly outperform shallow neural networks, even when normalized by computational cost.
Altogether, the results suggest that the neural network method can reliably produce solutions of acceptable accuracy for use in subsequent physical computations, such as particle simulations.
\end{abstract}

\maketitle

\section{Introduction}

Many important phenomena can be modelled effectively by partial differential equations (PDEs) with appropriate boundary conditions (BCs).
When PDE problems are posed in domains with complicated geometries, they are often too difficult to be solved analytically, and must instead be approximated numerically.
The standard tools for numerically solving PDE problems in complex geometries are mesh-based approaches, such as the finite element method (FEM) \cite{Cook2007}.
In these methods, the problem domain is decomposed into a mesh of smaller subdomains, and the solution is approximated by a linear combination of simple, local functions.

\begin{figure}
\includegraphics[width=0.8\columnwidth]{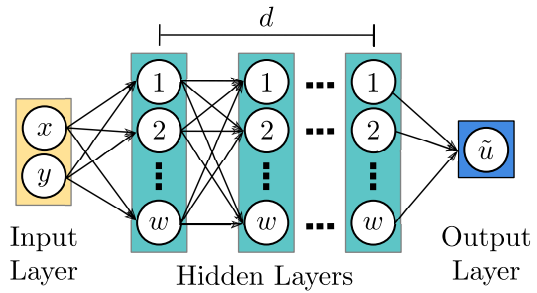}
\caption{Schematic of a fully-connected feedforward neural network of depth $d$ and width $w$ mapping coordinates $(x,y)$ to an output $\tilde{u}(x,y)$ . Each node computes a weighted sum of its incoming arrows, and the result (plus a bias) is passed to an activation function. In the NNM, the parameters are optimized to make $\tilde{u}(x,y)$ approximately satisfy a target PDE and its BCs.
\label{fig:nn-schematic}}
\end{figure}

In this work, we will explore a less common numerical solution method for PDE problems, which we will refer to as the neural network method (NNM) \cite{Dissanayake1994}.
In the NNM, the solution is directly approximated by a neural network (e.g., Fig~\ref{fig:nn-schematic}), rather than by a linear combination of local basis functions.
In a process called training, the network parameters are varied until it approximately satisfies the PDE and BCs.

The purpose of the present study is to investigate the effectiveness of the NNM on a problem exhibiting a complicated geometry.
Specifically, the NNM is used to solve a model of the electric field in the slit-well microfluidic device, which is an application of active research interest \cite{Han1999,Han2000,Levy2010,Dorfman2010}.
The problem domain is non-convex, and the electric field is discontinuous in the limit of sharp corners.
Despite the growing popularity of the NNM, relatively few authors have validated it on problems with such ill-behaved solutions.
The rest of this introduction provides an overview of the NNM, including its previous use to study systems similar to the slit-well, as well as a review of the slit-well device itself.

\subsection{The neural network method \label{sec:intro-nnm}}

The neural network method of solving differential equations was first published in 1994 by \citet{Dissanayake1994}, and belongs to the broader family of techniques known as methods of weighted residuals \cite{Dissanayake1994,Meade1994}.
Around the same time, \citet{MeadeJr1994} separately demonstrated a variant of the NNM that did not use iterative training, and instead solved a system of linear equations for the network weights; it was, however, designed for solving only ordinary differential equations.
In 1995, \citet{vanMilligen1995} independently proposed a method quite similar to the original approach by \citet{Dissanayake1994}, to solve second-order elliptic PDEs describing plasmas in tokamaks.
In 1998, the NNM was proposed independently again by \citet{Lagaris1998}.
Their modified methodology embedded the neural network within an ansatz that was manually constructed to exactly satisfy the boundary conditions; however, this form is challenging to construct when the boundary conditions or the domain geometry are complicated.
Many authors have since contributed to the development of the NNM, and in 2015 \citet{Yadav2015} published a book reviewing much of the work up to that time.

The NNM has various potential appeals over more common methods like FEM.
For instance, the NNM is mesh-free, and generally produces uniformly accurate solutions throughout the PDE domain \cite{Lagaris1997b,Yadav2015}.
Whereas earlier implementations used shallow neural networks (i.e.~those having only one hidden layer), many authors have recently noted the significant benefits of using deep architectures \cite{E2017,Avrutskiy2020,Berg2018,
Sirignano2018,Royo2017,Han2018,Magill2018neurips,
Hure2020,Karumuri2020,Mutuk2019,Nabian2019,Zhang2020,Beck2019}.
In particular, it appears that the NNM with deep neural networks performs remarkably well in high-dimensional problems \cite{E2017,Avrutskiy2020,
Sirignano2018,Royo2017,Han2018,
Wei2018,
Hure2020,Karumuri2020,Mutuk2019,Nabian2019,Zhang2020, Beck2019}.
Such high-dimensional PDEs are typically intractable using FEM and most traditional methods.
These suffer from the so-called curse of dimensionality, in which computational cost grows exponentially with the number of dimensions.
In addition to the above empirical demonstrations of the NNM, several theorems have been published stating that the computational cost of the NNM grows at most polynomially in the number of dimensions for various classes of PDEs \cite{Grohs2018,Jentzen2018,Hutzenthaler2019}.

Nonetheless, the theoretical grounding of the NNM is less thoroughly developed than those of other techniques.
There are as of yet few guarantees regarding, e.g., under what conditions the NNM will converge to the true solution of a given PDE, at what rate, and to what precision.
As such, confidence in the method still relies heavily on empirical demonstrations.
However, available empirical demonstrations focus primarily on problems with relatively well-behaved solutions \cite{Avrutskiy2020,
Royo2017,Han2018,
Hure2020,Karumuri2020,Mutuk2019,Zhang2020,Beck2019,McFall2009,
Berg2018,Nabian2019}.
Indeed, \citet{Michoski2020} noted this, and conducted an investigation of the NNM applied to irregular problems exhibiting shocks.
The current work is analogous in this regard, but focuses instead on the non-convexity of the slit-well domain as the source of irregularity.

\subsection{The slit-well microfluidic device \label{sec:intro-slitwell}}

\begin{figure}
\includegraphics[width=0.8\columnwidth]{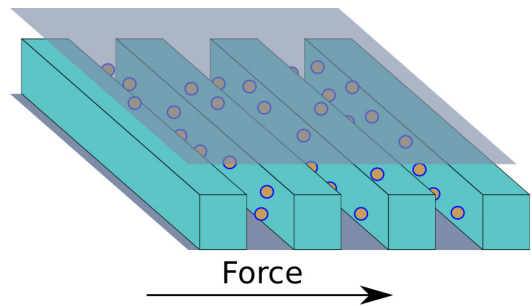}
\caption{A schematic of particles being electrically driven through the slit-well device. 
\label{fig:slit-well-schematic}}
\end{figure}

Micro- and nanofluidic devices (MNFDs) are small, synthetically fabricated systems with applications in molecular detection and manipulation \cite{Abgrall2008,Gardeniers2004,Mulero2010,Levy2010,Dorfman2010}.
One important use of MNFDs is to sort mixtures of molecules, including free-draining molecules such as DNA that cannot normally be separated electrophoretically in free solution \cite{Dorfman2010}.
For instance, the slit-well device proposed by \citet{Han1999} can be used for sorting polymers (such as DNA \cite{Han1999,Han2000,Fu2005,Fu2006}) or nanoparticles \cite{Cheng2008,Wang2019}.
The device's periodic geometry, illustrated schematically in Fig.~\ref{fig:slit-well-schematic}, consists of parallel channels (called wells) separated by shallower regions (called slits).
An electric field is applied to drive molecules through the device.

MNFDs such as the slit-well exploit the complexity of physical phenomena at the single-molecular scale (often below the optical resolution limit) to produce useful and sometimes surprising behaviours.
This, however, makes them challenging to design and optimize, and renders theoretical and computational investigations important to the development of MNFD technologies.
For example, the sorting mechanism in the slit-well device depends nonlinearly on the magnitude of the applied electric field as well as the size and shape of the wells, the slits, and the molecules themselves \cite{Cheng2008,Fu2005,Fu2006,Dorfman2010,Langecker2011,Wang2019}.
For some choices of these parameters, the slit-well sorts molecular mixtures into increasing order of size; for others, however, it sorts them into decreasing order.
A rich literature exists exploring these processes, reviewed in part by \citet{Dorfman2010} and \citet{Langecker2011}.

\subsection{NNM with complicated geometries \label{sec:intro-geo}}

There are relatively few demonstrations of the NNM on problems with complicated domain geometries.
Specifically, the NNM has mostly been applied to problems posed in rectangular or circular domains \cite{Avrutskiy2020,
Royo2017,Han2018,
Hure2020,Karumuri2020,Mutuk2019,Zhang2020,Beck2019}.
Of note, \citet{Wei2018} used the NNM to solve PDEs in nanobiophysics that also arise in MNFDs (i.e., Fokker-Planck for particles and polymers).
However, their work did not consider these problems in MNFD geometries.
Even among the demonstrations of the NNM in more complicated (e.g., non-convex) domain geometries, most problems feature boundary conditions that produce relatively smooth, well-behaved solutions \cite{McFall2009,
Berg2018,Nabian2019}.
\citet{Sirignano2018} solved a free-boundary problem based on a financial system, but it is not clear whether that PDE exhibits the specific kinds of challenging features considered in the current work.

An exception to the above is given by \citet{E2017}, who applied a variant of the NNM to a Poisson equation in a square domain with a re-entrant needle-like boundary.
This problem exhibits the same singular behaviour as the slit-well problem with sharp corners (see Sec.~\ref{sec:method-problem}). 
Their Deep Ritz training protocol was based on a variational formulation of Poisson's equation.
However, variational formulations cannot be obtained for all PDEs \cite{Finlayson2013}.
For this reason, we have opted to study the more general NNM algorithm originally presented by \citet{Dissanayake1994}.

When \citet{Anitescu2019} revisited this needle problem using the original method of \citet{Dissanayake1994}, they reported poorer convergence than obtained by \citet{E2017} with the Deep Ritz method.
A similar observation was made during the present work: re-entrant corners significantly impair the convergence of the standard NNM (Sec.~\ref{sec:method-problem}).
In contrast to the present work, the error analyses reported by \citet{E2017} and \citet{Anitescu2019} did not consider the physical realism of the NNM solutions (Sec.~\ref{sec:intro-realism}) nor the accuracy of the NNM solutions' gradients.
These characteristics of the NNM are important for use in various applications, including studies of MNFDs, and are investigated directly in the present work.

\subsection{Physical realism of NNM solutions \label{sec:intro-realism}}

Various modifications of the NNM have been proposed to ensure solutions exactly satisfy problem-specific invariants that are known \textit{a priori}, such as boundary conditions \cite{Lagaris1997b, McFall2009, Berg2018}, non-negativity \cite{AlAradi2018}, Hamiltonian dynamics \cite{Mattheakis2019}, or special invariants of the Schr\"{o}dinger equation \cite{Hermann2019}.
However, manually creating formulations of the NNM that explicitly satisfy specific invariants can be difficult.
Furthermore, this approach cannot account for invariants which may be unknown ahead of time.
It is natural to question how well the NNM approximates invariant quantities when these are not explicitly enforced.

In fact, although certain numerical methods can be devised specifically to satisfy some conservation laws (e.g., finite volume methods conserve flux \cite{Moukalled2016}, symplectic ODE integrators conserve energy \cite{Hairer2006}), most numerical methods (including standard FEM formulations) do not satisfy physical invariances exactly.
For instance, \citet{Zhang2017} discussed what modifications of the FEM are necessary to render it flux-conserving.
As part of the present work, we will investigate how well the NNM satisfies physical invariances of the slit-well problem in the absence of any problem-specific customization.

\section{Methodology \label{sec:method}}

\subsection{Problem statement \label{sec:method-problem}}

\begin{figure}
\includegraphics[width=0.8\columnwidth]{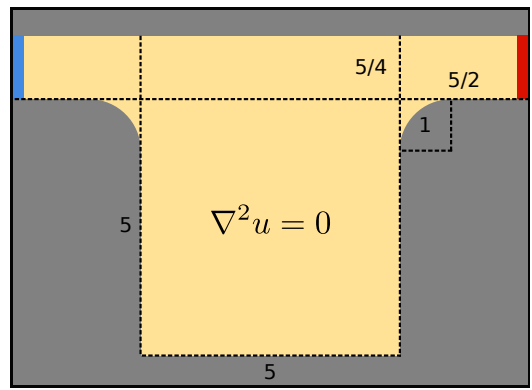}
\caption{A cross-sectional view of the slit-well device illustrating our PDE model of the electric potential in one periodic sub-unit of the device. The re-entrant corners follow circular arcs, and the numbers indicate the lengths of each dotted line. The solution satisfies Laplace's equation in the yellow region, Dirichlet conditions on the red and blue boundaries, and homogeneous Neumann conditions on the gray boundaries.
\label{fig:slit-well-pde}
}
\end{figure}

We use the simplest electrostatic model of the electric field $\mathbf{E}$ in the slit-well, namely the two-dimensional Laplace equation for the electric potential $u$.
Figure~\ref{fig:slit-well-pde} illustrates the geometry of our model over one periodic sub-unit of the slit-well device.
Uniform Dirichlet boundary conditions were imposed on the coloured segments (specifically, $u=\pm1$ on the right and left, respectively) to model an applied voltage across the system.
The grey boundaries correspond to homogeneous Neumann (i.e., insulating) boundary conditions.
Throughout the interior of the domain (i.e., the yellow area in Fig.~\ref{fig:slit-well-pde}), the potential was modelled by Laplace's equation.

In contrast with other authors, we have rounded the re-entrant corners at the interface of the slits and wells.
It can be shown that near sharp (i.e. non-differentiable) re-entrant corners, solutions $u$ to Laplace's equation are not continuously differentiable \cite{Cai2001, Dauge2006, Grisvard1985}.
That is, sharp re-entrant corners cause singularities in the electric field $\mathbf{E}$.
Because the magnitude of $\mathbf{E}$ near the corners diverges as the curvature goes to zero, the slit-well electric field is ill-conditioned, in the sense that small changes in the curvature of the corners produce large changes in $\mathbf{E}$.

Although such ill-conditioning hinders the performance of most numerical methods, including FEM
\cite{Cai2001, Dauge2006, Grisvard1985}, they present a particular challenge for the NNM.
The fully-connected feedforward neural networks typically used for the NNM are infinitely differentiable functions.
However, the true solution to the slit-well problem with sharp corners exhibits a discontinuous electric field, so that significant errors seem likely near the corners.
Furthermore, because the neural network is a global approximation method, local errors near the corners can affect performance throughout the domain.

In practice, the training methodology we present here (Sec.~\ref{sec:method-nnm}), when applied to the problem with sharp corners, failed to converge to even a reasonable approximation of the true solution.
Even in preliminary tests with rounded corners, the convergence rate of the NNM was observed to deteriorate as the curvature of the corners was reduced.
Therefore, for the current work, an intermediate curvature (Fig.~\ref{fig:slit-well-pde}) was selected to produce a challenging but attainable benchmark for the NNM.

\subsection{NNM implementation \label{sec:method-nnm}}

In this section, we describe our implementation of the NNM.
It is similar to those of \citet{Dissanayake1994,vanMilligen1995,
Berg2018,Sirignano2018,Magill2018neurips}, and \citet{Wei2018}, among others.
The true solution $u(\mathbf{x})$ to the PDE was directly approximated by a neural network $\tilde{u}(\mathbf{x})$.
This was accomplished by training the neural network to minimize the loss functional
\begin{align}
\mathcal{L}[\tilde{u}] = \int_\Omega \left( \nabla^2 \tilde{u} \right)^2 dA + \int_{\partial \Omega} \left( B[\tilde{u}] \right)^2 ds. \label{eqn:loss}
\end{align}
Here $\nabla^2 u = 0$ is the PDE required to hold in the interior of the domain $\Omega \subset \mathbb{R}^2$, and $B$ is a differential operator describing the boundary conditions $Bu=0$ on the boundary $\partial \Omega$ of the domain (described in Section~\ref{sec:method-problem} and illustrated in Fig.~\ref{fig:slit-well-pde}).
Thus, $\mathcal{L}[\tilde{u}]$ quantifies the extent to which the neural network fails to satisfy the PDE and its boundary conditions.

The parameters of the network were updated iteratively using the Adam optimizer, a modified gradient descent algorithm \cite{Kingma2014}.
The integrals in $\mathcal{L}[\tilde{u}]$ were approximated via the Monte Carlo method, as described in more detail below.
The approximate electric field, $\mathbf{\tilde{E}}$, and other required derivatives were obtained exactly via automatic differentiation.
The weights of the network were initialized by the Glorot method \cite{Glorot2010}.
Computations were done using Tensorflow 1.13, and all hyperparameters not discussed here were set to their default values \cite{Tensorflow2015}.

The Monte Carlo samples $\mathbf{x}_i \in \Omega$ used to estimate the first term of $\mathcal{L}[\tilde{u}]$ were selected from 100,000 points uniformly distributed in the bounding rectangle $[-L_x,L_x]\times[-L_y,L_y]$, by rejecting those lying outside the domain.
Those used to estimate the second term were generated by directly sampling the boundary with a linear density of 40 points per unit length.
Altogether, this yielded an expected batch size of roughly 62,000.
To reduce the overhead of sampling training points,  batches were re-used for 1,000 parameter updates before resampling.

The testing loss was computed on a set of points sampled once at the beginning of training, generated using the same procedure as the training points.
The testing loss was computed and recorded every 100 parameter updates.
Early stopping was used to terminate training when the testing loss failed to improve after 100 consecutive tests.
The final network was taken from the epoch at which the testing loss was smallest. 
This training procedure was conceived to ensure that networks converged to very stable local minima, in order to study the behaviour of the NNM in the limit of long training time.

The neural networks considered in this study were all fully-connected feedforward networks with $\tanh$ activation functions (Fig.~\ref{fig:nn-schematic}), consisting of $d$ hidden layers of equal width $w$.
Specifically, the networks mapped an input vector $\mathbf{x}$, corresponding to a point in the problem domain, to $\tilde{u}$ given by
\begin{align}
\tilde{u}(\mathbf{x}) = f_{d+1} \circ f_d \circ \cdots \circ f_1 (\mathbf{x}),
\end{align}
where
\begin{align}
f_1(\mathbf{x}) &= \tanh\left( W_1 \mathbf{x} + \mathbf{b}_1 \right), \\
f_i(\mathbf{x}) &= \tanh\left( W_i f_{i-1}(\mathbf{x}) + \mathbf{b}_i \right), & i=2\ldots d, \\
f_{d+1}(\mathbf{x}) &= \textbf{W}_{d+1} f_d(\mathbf{x}) + b_{d+1}.
\end{align}
Here, $W_1 \in \mathbb{R}^{w\times 2}$, $W_i \in \mathbb{R}^{w\times w}$ for $i = 2\ldots d$, and $W_{d+1} \in \mathbb{R}^{1\times w}$ are the network's weight matrices, while $\mathbf{b}_i \in \mathbb{R}^{w}$ for $i = 1\ldots d$, and $b_{d+1} \in \mathbb{R}$ are its biases.

\subsection{FEM implementation \label{sec:method-fem}}

To provide a reliable ground truth against which to compare the performance of the NNM, the target PDE was also solved via the FEM using FEniCS \cite{Fenics2015}.
The domain and mesh were constructed using the mshr package.
The resolution parameter for \verb-generate_mesh- was set to 200 and the circular re-entrant corners were approximated linearly with 100 segments each.

In order to obtain an accurate approximation of the electric field, and not just of the electric potential, the FEM was applied to a standard dual-mixed formulation of Laplace's equation for the electric field and electric potential simultaneously \cite{Fenics2015}.
In this approach, $\tilde{u}$ and $\mathbf{\tilde{E}}$ are approximated simultaneously using separate basis functions.
Solving for $\tilde{u}$ alone and reconstructing $\mathbf{\tilde{E}}$ by differentiation was found to yield poor results.

Convergence tests (not shown) confirmed that the FEM solution converged in proportion to the square of the mesh resolution.
The tests suggest that the absolute error in the FEM solution relative to the true solution is on the order of machine precision (i.e, $10^{-16}$).
Note that the FEM solution was computed in double precision, whereas the NNM was computed in single precision.

\section{Results \label{sec:results}}

\begin{figure*}
\includegraphics[width=\textwidth]{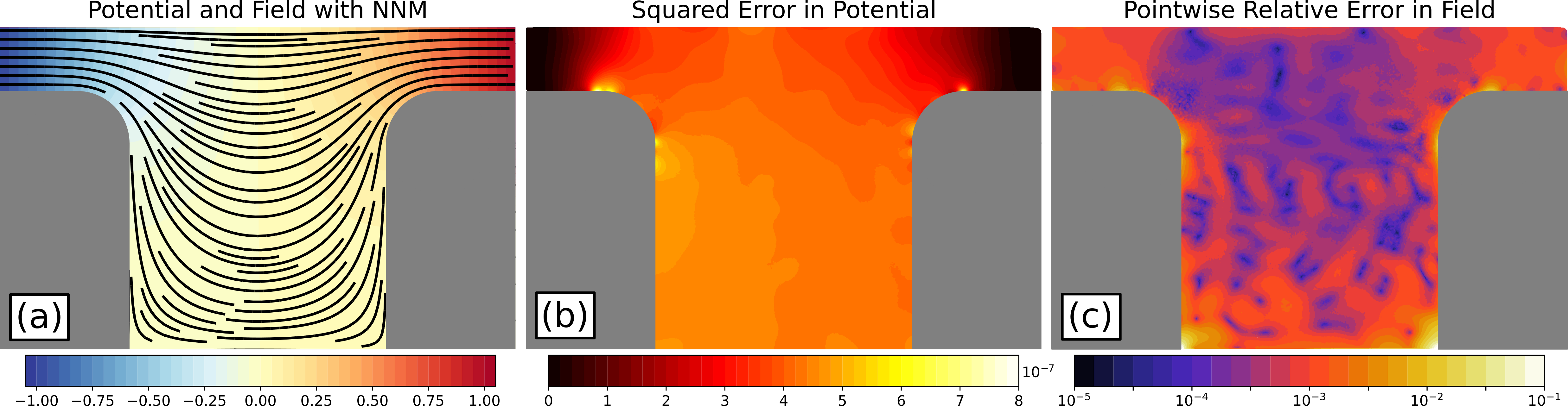}
\caption{Example NNM solution using 5 hidden layers of width 75. (a) Approximate electric potential (colored contours) and electric field (solid black lines). (b) Squared error of the electric potential. (c) Pointwise relative error in the electric field (note the logarithmic color scale).
The error in plots (b) and (c) are interpolated from values evaluated on the FEM mesh points. \label{fig:slit-well-soln}
}
\end{figure*}

At its core, the NNM is motivated by the rationale that training networks to minimize the loss functional (Eqn.~\ref{eqn:loss}) will cause those networks to approximate the correct solution.
This section contains investigations into the following related questions:
\begin{enumerate}
\item If a network exhibits a small loss, how close is it to the true solution? Specifically, is the loss functional a reliable estimator of actual network performance?
\item If a network is close to the true solution, how well does it reproduce the physical characteristics of the true solution? Specifically:
\begin{itemize}
\item[-] To what extent does it exhibit the same spatial symmetries as the true solution?
\item[-] To what extent does it conserve electric flux?
\end{itemize}
\item If a network is close to the true solution, and the corresponding electric field is used to conduct particle simulations, how accurate are subsequent measurements made using those particle simulations?
\item How does architecture affect these conclusions?
\end{enumerate}
All experiments were repeated across four random initializations and multiple network architectures: specifically, all combinations of depths $d=1,2,3,4,5,6$ and widths $w=10,25,50,75,100,150,200,250$ were examined, as well as networks of depth 1 and widths 500 and 1000.

Figure~\ref{fig:slit-well-soln}(a) shows an example of an NNM solution obtained using a network of depth 5 and width 75.
The approximate electric field $\mathbf{\tilde{E}}$ is superimposed in black lines over colored contours showing the approximate electric potential $\tilde{u}$.
It is visually indistinguishable from the reference FEM solution (not shown).
Figure \ref{fig:slit-well-soln}(b) shows $(\tilde{u}-u)^2$, the squared error of the NNM potential compared to the FEM potential.
Figure \ref{fig:slit-well-soln}(c) shows $\Vert \mathbf{\tilde{E}} - \mathbf{E} \Vert_2 / \Vert \mathbf{E} \Vert_2$, the pointwise relative error of the NNM electric field.
Here, $\| \cdot \|_2$ denotes the Euclidean norm.
Note that the error in the potential cannot be normalized pointwise, as discussed in the next section.

Both of the error distributions in Fig.~\ref{fig:slit-well-soln} are particularly pronounced near the re-entrant corners.
The electric field intensity is also very large in these regions (see Fig.~\ref{fig:slit-well-absE}).
In the limit of small curvature, in fact, it is at these corners that the electric field develops singularities (see Sec.~\ref{sec:method-problem}).
In fact, the peaks in error and electric field intensity both occur precisely where the boundary transitions from flat to curved, i.e., where the second derivative of the boundary curve is discontinuous.

Additionally, Fig.~\ref{fig:slit-well-soln}(c) shows pronounced relative error in the electric field near the corners at the bottom of the well.
These peaks arise because the magnitude of the true electric field approaches zero in those corners (see Fig.~\ref{fig:slit-well-absE}).
Since the denominator of $\Vert \mathbf{\tilde{E}} - \mathbf{E} \Vert_2 / \Vert \mathbf{E} \Vert_2$ is very small, even small errors in the electric field near those corners manifest as large relative error.
The maximum relative error in the domain $\Omega$ consistently occured in these bottom-most corners for all NNM solutions in the dataset.
Nonetheless, for many applications, errors of this kind are likely to be less important than the errors occuring near the re-entrant corners, as they are much smaller in absolute magnitude.

\subsection{Error relative to FEM \label{sec:results-standard}}

\begin{figure}
\includegraphics[width=\columnwidth]{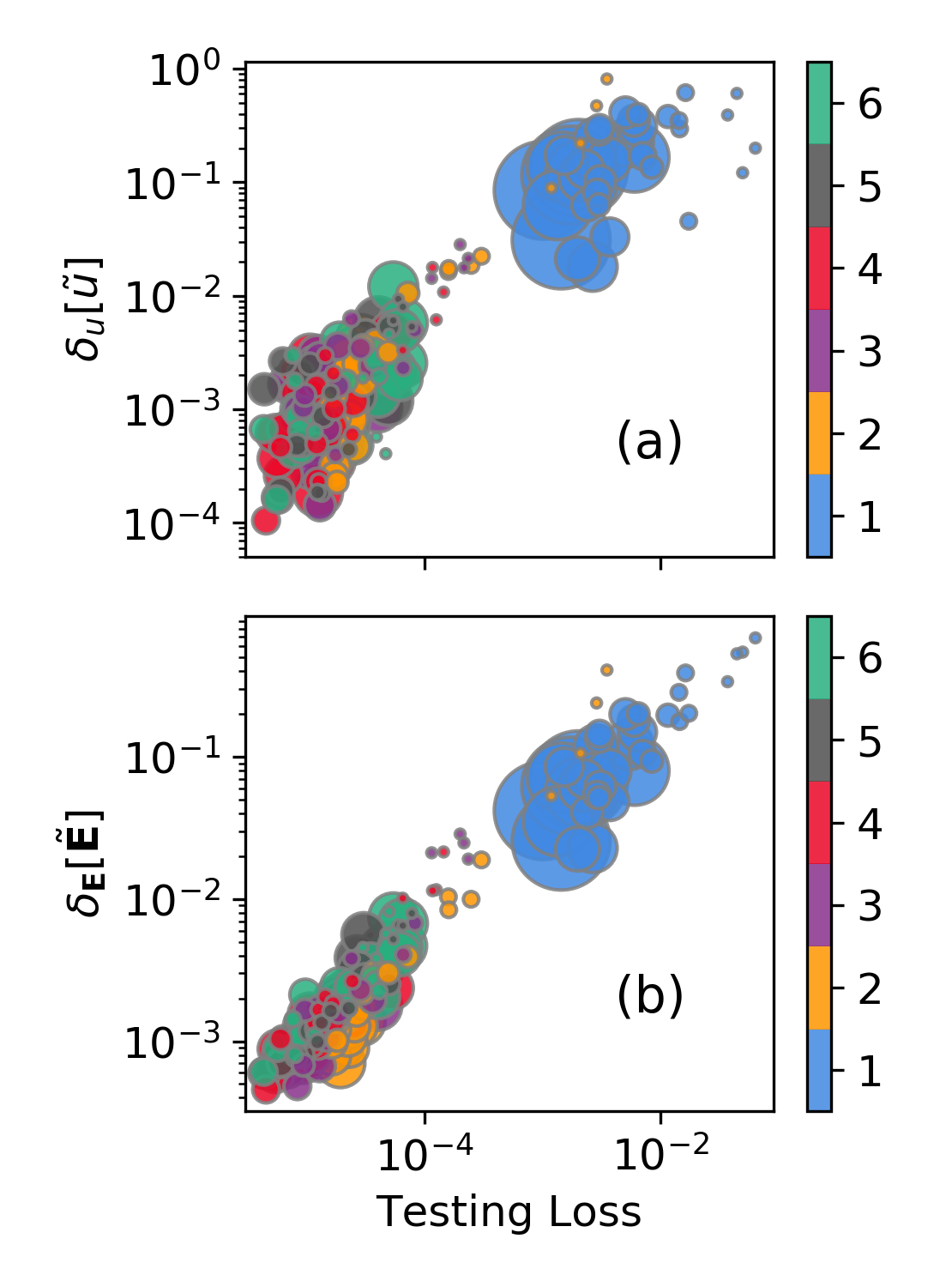}
\caption{Global error metrics for the NNM solutions relative to the reference FEM solution, shown against testing loss for a variety of network architectures. 
(a) The relative error of the electric potentials, $\delta_u[\tilde{u}]$. (b) The relative error of the electric fields, $\delta_{\mathbf{E}}[\mathbf{\tilde{E}}]$.
Marker colour indicates the depth of the network, and marker area indicates its width. 
\label{fig:err_v_loss}}
\end{figure}

The purpose of this section is to investigate the errors of the NNM solutions relative to the reference FEM solution, and to what extent the loss functional correlates with these errors.
The error in an approximate electric potential $\tilde{u}$ will be characterized by
\begin{align}
\delta_u[\tilde{u}] = \sqrt{\frac{\left\langle \left(\tilde{u} - u\right)^2 \right\rangle_\Omega}{\left\langle u^2 \right\rangle_\Omega}}.
\end{align}
Here, $\langle \cdot \rangle_\Omega$ denotes the mean over the domain $\Omega$.
Whereas Fig.~\ref{fig:slit-well-soln}(b) shows the distribution of the squared error in $\tilde{u}$ throughout the domain, $\delta_u[\tilde{u}]$ corresponds to the root mean squared error of $\tilde{u}$ over $\Omega$, normalized by the root mean squared value of the true solution $u$.
Note that one cannot define an unambiguous pointwise relative error for $\tilde{u}$, since the electric potential does not have a physically meaningful zero.
The metric $\delta_u[\tilde{u}]$ represents the magnitude of the error in $\tilde{u}$ relative to the magnitude of the true solution $u$, when both of these are measured in the $L^2$ norm for functions.

For the electric field, conversely, a meaningful pointwise relative error can be defined as $\Vert \mathbf{\tilde{E}} - \mathbf{E} \Vert_2 / \Vert \mathbf{E} \Vert_2$, where both the numerator and the denominator vary throughout the domain.
The average of this pointwise relative error is denoted
\begin{align}
\delta_{\mathbf{E}}[\mathbf{\tilde{E}}] = \left\langle \frac{\Vert \mathbf{\tilde{E}} - \mathbf{E} \Vert_2 }{ \Vert \mathbf{E} \Vert_2 } \right\rangle_\Omega,
\end{align}
and acts as a global error metric for $\mathbf{\tilde{E}}$.
This is precisely the mean of error distributions like the one shown in Fig.~\ref{fig:slit-well-soln}(c).

Figure~\ref{fig:err_v_loss} shows the global error metrics $\delta_u[\tilde{u}]$ and $\delta_{\mathbf{E}}[\mathbf{\tilde{E}}]$ for all networks in the dataset, plotted against each network's testing loss.
The integrals required to compute the error metrics were approximated via the Monte Carlo method, by sampling the domain interior using the same procedure described in Section.~\ref{sec:method-nnm}.
Marker color corresponds to network depth, and marker size corresponds to network width.

It is clear in Fig.~\ref{fig:err_v_loss} that lower testing losses correlate strongly with lower values of both $\delta_u[\tilde{u}]$ and $\delta_{\mathbf{E}}[\mathbf{\tilde{E}}]$.
This result confirms the basic motivation underlying the NNM, namely that training neural networks to minimize the loss functional will cause them to approximate the correct solution.
It also suggests that, in the absence of theoretical guarantees on the convergence of the NNM, the testing loss may provide a practical proxy for estimating a solution's true accuracy.

The data in both Figs.~\ref{fig:err_v_loss}(a) and (b) partition conveniently into two clusters.
The upper-right clusters consist of those networks achieving relative errors worse than 1\% in both $\delta_u[\tilde{u}]$ and $\delta_{\mathbf{E}}[\mathbf{\tilde{E}}]$.
This population contains all of the shallow network architectures, suggesting that at least two hidden layers are required to achieve good performance on this problem.
Furthermore, as discussed in Sec.~\ref{sec:results-arch}, shallow networks always underperform relative to deep networks, even when normalized by capacity.
The narrowest of the deep network architectures also attain relative errors worse than 1\%.
This implies that even with two hidden layers, networks require some minimum capacity (i.e., memory consumption) in order to achieve good performance on this problem.

The lower-left clusters in Figs.~\ref{fig:err_v_loss}(a) and (b) contain the majority of the dataset, and consist of those networks attaining relative errors below 1\% in both $\delta_u[\tilde{u}]$ and $\delta_{\mathbf{E}}[\mathbf{\tilde{E}}]$.
The best networks achieved relative errors as low as $\delta_u[\tilde{u}] \approx 0.01\%$ and $\delta_{\mathbf{E}}[\mathbf{\tilde{E}}] \approx 0.1\%$.
For reference, the example solution shown in Fig.~\ref{fig:slit-well-soln} corresponds to a testing loss of $\mathcal{L}[\tilde{u}] \approx 9 \cdot 10^{-6}$, and error values of $\delta_u[\tilde{u}] \approx 0.2\%$ and $\delta_{\mathbf{E}}[\mathbf{\tilde{E}}] \approx 0.08\%$.
A variety of architecture choices (i.e., depths and widths) produce comparably good performance, suggesting that the NNM can produce accurate solutions without the need for careful architecture tuning.
This is explored further in Sec.~\ref{sec:results-arch}.

\subsection{Physically motivated error metrics \label{sec:results-special}}

The results in the previous section suggest that the NNM can reliably produce accurate solutions to the slit-well problem.
Furthermore, networks with smaller loss values are closer to the true solution, i.e., they have smaller error values.
Finally, the NNM does not appear overly sensitive to the choice of architecture, given at least two hidden layers and sufficient network width.

The purpose of this section is to investigate whether networks with small loss and error values also approximately reproduce physical characteristics of the true solution.
Specifically, we investigate the NNM solutions' satisfaction of spatial symmetries and the conservation of electric flux.

\subsubsection{Deviation from symmetry \label{sec:results-special-symm}}

The true solution of the target PDE satisfies three spatial symmetries.
First, the true electric potential $u$ is anti-symmetric in the horizontal direction about the centre of the well, i.e.,
\begin{align}
u(x,y) = -u(-x,y),
\end{align}
where $(x,y)$ are the coordinates of a point about the center of the well.
As a result, the vertical component of the true electric field $\mathbf{E}$ also exhibits this anti-symmetry in $x$, i.e.,
\begin{align}
E_y(x,y) = -E_y(-x,y).
\end{align}
Finally, the horizontal component of the electric field is symmetric about the centre of the domain, i.e.,
\begin{align}
E_x(x,y) = E_x(-x,y).
\end{align}

The extent to which a network deviates from these symmetries will be quantified using relative error metrics analogous to those used in the previous section.
Specifically, the deviation of an approximate electric potential $\tilde{u}$ from symmetry will be quantified by
\begin{align}
R_u[\tilde{u}] = \sqrt{ \frac{ \left\langle \left( \tilde{u} - \tilde{u}' \right)^2 \right\rangle_\Omega }{ \left\langle u^2 \right\rangle_\Omega } },
\end{align}
where $\tilde{u}'(x,y) = -\tilde{u}(-x,y)$.
This is the root mean squared difference between $\tilde{u}$ and its negative reflection, normalized by the root mean squared value of the true potential $u$.
In analogy with $\delta_u[\tilde{u}]$, the metric $R_u[\tilde{u}]$ measures the magnitude of the deviation of $\tilde{u}$ from symmetry relative to the magnitude of the true solution $u$ (when both are measured in the $L^2$ norm).
The deviation of an approximate electric field $\mathbf{\tilde{E}}$ from symmetry will be quantified by
\begin{align}
R_{\mathbf{E}}[\mathbf{\tilde{E}}] = \left\langle \frac{\Vert \mathbf{\tilde{E}} - \mathbf{\tilde{E}}' \Vert_2 }{ \Vert \mathbf{E} \Vert_2 } \right\rangle_\Omega,
\end{align}
where $\mathbf{\tilde{E}}'$ is the transformed electric field
\begin{align}
\tilde{E}_x'(x,y) &= \tilde{E}_x(-x,y) \\
\tilde{E}_y'(x,y) &= -\tilde{E}_y(-x,y).
\end{align}
In analogy with $\delta_{\mathbf{E}}[\mathbf{\tilde{E}}]$, this is the mean pointwise relative deviation from symmetry of the electric field.

These metrics of deviation from symmetry are closely connected to the relative error metrics of Sec.~\ref{sec:results-standard}.
Specifically, the triangle inequality implies that
\begin{align}
\sqrt{\left\langle \left( \tilde{u} - \tilde{u}' \right)^2 \right\rangle_\Omega} \leq \sqrt{\left\langle \left( \tilde{u} - u \right)^2 \right\rangle_\Omega} + \sqrt{\left\langle \left( u - \tilde{u}' \right)^2 \right\rangle_\Omega}.
\end{align}
By definition, the true solution $u$ is invariant under the transformation that maps $\tilde{u}$ to $\tilde{u}'$.
Specifically,
\begin{align}
\tilde{u}(x,y) - u(x,y) = -\tilde{u}'(-x,y) - \left( -u(-x,y) \right).
\end{align}
By the symmetry of the domain, it follows that
\begin{align}
\sqrt{\left\langle \left( \tilde{u} - u \right)^2 \right\rangle_\Omega} = \sqrt{\left\langle \left( u - \tilde{u}' \right)^2 \right\rangle_\Omega}.
\end{align}
Combining these results and dividing by $\sqrt{\left\langle u^2 \right\rangle_\Omega}$, it follows that
\begin{align}
R_u[\tilde{u}] \leq 2 \delta_u[\tilde{u}], \label{eqn:Ru-leq-du}
\end{align}
that is, the distance from an approximate potential $\tilde{u}$ to its reflection $\tilde{u}'$ is, at most, twice the distance from $\tilde{u}$ to the true solution $u$.
Very similar reasoning can be applied to an approximate electric field $\mathbf{\tilde{E}}$ to conclude that
\begin{align}
R_{\mathbf{E}}[\mathbf{\tilde{E}}] \leq 2 \delta_{\mathbf{E}}[\mathbf{\tilde{E}}]. \label{eqn:RE-leq-dE}
\end{align}

Thus, solutions with small error values will inevitably be nearly symmetric, simply by virtue of being nearly equal to a symmetric function.
Furthermore, since it was established in Sec.~\ref{sec:results-standard} that the loss functional provides a reliable estimator of the error, it follows that the loss also provides a reliable estimator of the deviation from symmetry.
It remains to be seen, however, whether or not inequalities \ref{eqn:Ru-leq-du} and \ref{eqn:RE-leq-dE} are strict in practice.
That is, do neural networks learn that symmetry is a desirable feature, or are they only symmetric insofar as they approximate the true solution?

\begin{figure}
\includegraphics[width=\columnwidth]{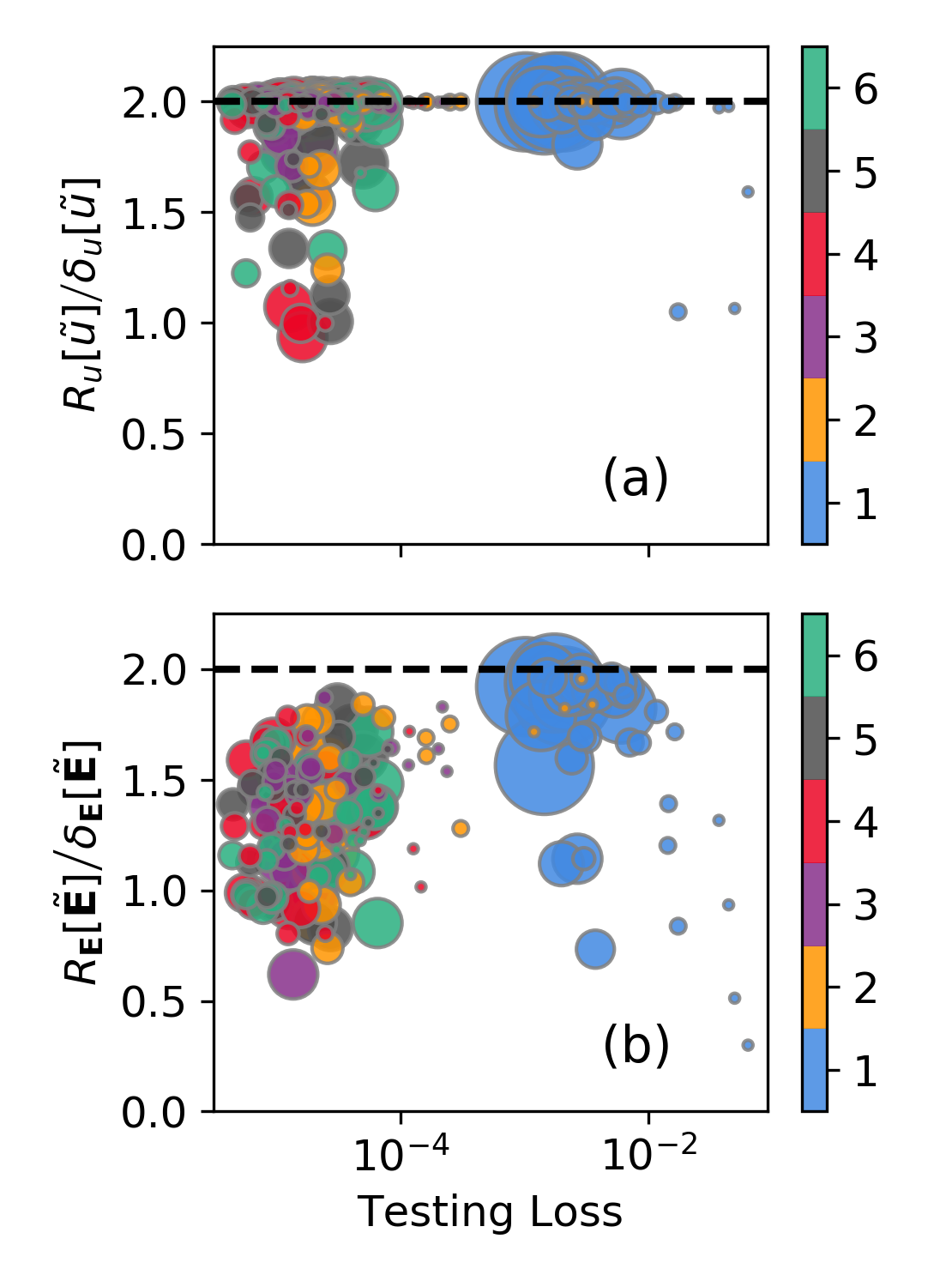}
\caption{Relative deviation of symmetry for the NNM solutions normalized by relative error, shown against testing loss.
(a) Deviation of symmetry of the NNM electric potentials, $R_u[\tilde{u}]$, divided by the relative error, $\delta_u[\tilde{u}]$.
(b) Deviation of symmetry of the NNM electric fields, $R_{\mathbf{E}}[\mathbf{\tilde{E}}]$, divided by the relative error, $\delta_{\mathbf{E}}[\mathbf{\tilde{E}}]$. 
Marker colour indicates the depth of the network, and marker area indicates its width. 
The dotted lines show the upper bounds given by Equations \ref{eqn:Ru-leq-du} and \ref{eqn:RE-leq-dE}.
 \label{fig:sym_v_loss}}
\end{figure}

Figure~\ref{fig:sym_v_loss} shows $R_u[\tilde{u}]/\delta_u[\tilde{u}]$ and  $R_{\textbf{E}}[\mathbf{\tilde{E}}]/\delta_{\mathbf{E}}[\mathbf{\tilde{E}}]$ for all networks in the dataset, plotted against each network's testing loss.
As in Fig.~\ref{fig:err_v_loss}, the marker sizes correspond to network widths, and the colours indicate network depth.
The dotted lines correspond to the maximum deviation from symmetry permitted for a given error value, according to inequalities \ref{eqn:Ru-leq-du} and \ref{eqn:RE-leq-dE}.

Most of the data in Fig.~\ref{fig:sym_v_loss}(a) lie nearly on the dotted line: roughly 90\% lie above 1.5, and 75\% lie above 1.9.
This indicates that most of the electric potentials approximated via the NNM satisfy the target symmetries only to the smallest degree required by virtue of their proximity to the true solution.
The data in Fig.~\ref{fig:sym_v_loss}(b), however, lie somewhat farther from the dotted line.
Quite a few of the most symmetric electric field approximations have $R_{\textbf{E}}[\mathbf{\tilde{E}}]/\delta_{\mathbf{E}}[\mathbf{\tilde{E}}]$ ratios below 1, indicating that they are more similar to their own reflections than they are to the true solution.
It is important to note, however, that the electric field metrics of error and symmetry are normalized pointwise by the electric field intensity, whereas the electric potential metrics are not normalized pointwise.
This distinction may account for some of the apparent differences between Figs.~\ref{fig:sym_v_loss}(a) and (b).

Altogether, the results in this section indicate that the NNM solutions deviate from the symmetries of the true solution by an amount comparable to their error values.
Some networks may produce electric field solutions that are more symmetric than required given their error values alone, but most networks only exhibit the minimal degree of symmetry required by the triangle inequality.
As discussed in the introduction, directly constraining the networks to satisfy the symmetries (e.g., by modifying the network architectures, or by adding additional terms to the loss functional) would almost certainly improve the symmetry of the resulting approximations.
However, implementing such constraints can be expensive for more complicated invariants, and some problems may exhibit invariants that are unknown \textit{a priori}.
These results illustrate that the NNM can still learn to satisfy invariants approximately, even when they are not explicitly enforced.
Furthermore, the loss functional may provide a means of empirically estimating the extent to which such invariants are satisfied in practice.

\subsubsection{Conservation of flux \label{sec:results-special-flux}}

Another important physical property of the true solution to the target PDE is the conservation of electric flux.
In its strong form, conservation states that the true electric field $\mathbf{E}$ must be divergence-free at all points in the domain.
This is equivalent to the condition that the true electric potential $u$ must satisfy Laplace's equation, $\nabla^2 u = 0$, since it can be rewritten as
\begin{align}
\nabla \cdot (\nabla u) = \nabla \cdot \mathbf{E} = 0.
\end{align}
Thus, one could quantify the deviation from conservation of flux of an approximate field $\mathbf{\tilde{E}}$ by computing some error norm of $\nabla \cdot \mathbf{\tilde{E}}$.
However, since all the derivatives taken in the NNM are exact (obtained via automatic differentiation), $\nabla \cdot \mathbf{\tilde{E}}$ is exactly equal to $\nabla^2 \tilde{u}$.
As a result, the first term of the loss functional (Eqn.~\ref{eqn:loss}) is precisely a measure of how well the NNM satisfies the strong form of the conservation of flux.

Nonetheless, the strong form of conservation is insufficient to fully describe the extent to which the electric field conserves flux over extended regions of space within the domain.
This is better described using the weak form, which states that the surface integral of the flux into any closed subset of the domain must be zero.
Motivated by this, we define the quantity
\begin{align}
\mathcal{E}(\tilde{u}; \epsilon) = \frac{1}{|\Omega^\epsilon|} \int_{\Omega^\epsilon} \left[ \frac{1}{|B_{\epsilon}|} \int_{\partial B(\mathbf{x};\epsilon)} \tilde{E}_{\hat{n}} ds \right]^2 dA. \label{eqn:field-based-flux-loss}
\end{align}
Here $B(\mathbf{x};\epsilon)$ is a ball of radius $\epsilon$ centered at a point $\mathbf{x}$ in the domain, $\partial B(\mathbf{x};\epsilon)$ denotes its boundary, and $\tilde{E}_{\hat{n}}$ denotes the outward normal component of the electric field into its surface.
The outer integral is taken over $\Omega^\epsilon$, by which we denote the set of all points in the domain that are at least a distance $\epsilon$ from the boundary.
The factors $|\Omega^\epsilon|$ and $|B_{\epsilon}|$ are the areas of $\Omega^\epsilon$ and $B(\mathbf{x};\epsilon)$, respectively.
In other words, $\mathcal{E}(\tilde{u}; \epsilon)$ is the mean square norm of the flux into all balls of radius $\epsilon$ that are entirely contained within $\Omega$, divided by the area of those balls.
Because this definition of $\mathcal{E}(\tilde{u}; \epsilon)$ is mesh-agnostic, it can also be computed directly for a FEM solution.
Numerical calculations of $\mathcal{E}(\tilde{u}; \epsilon)$ and related metrics in this section are somewhat technical, and details are relegated to App.~\ref{app:fluxcalc}.

\begin{figure}
\includegraphics[width=\columnwidth]{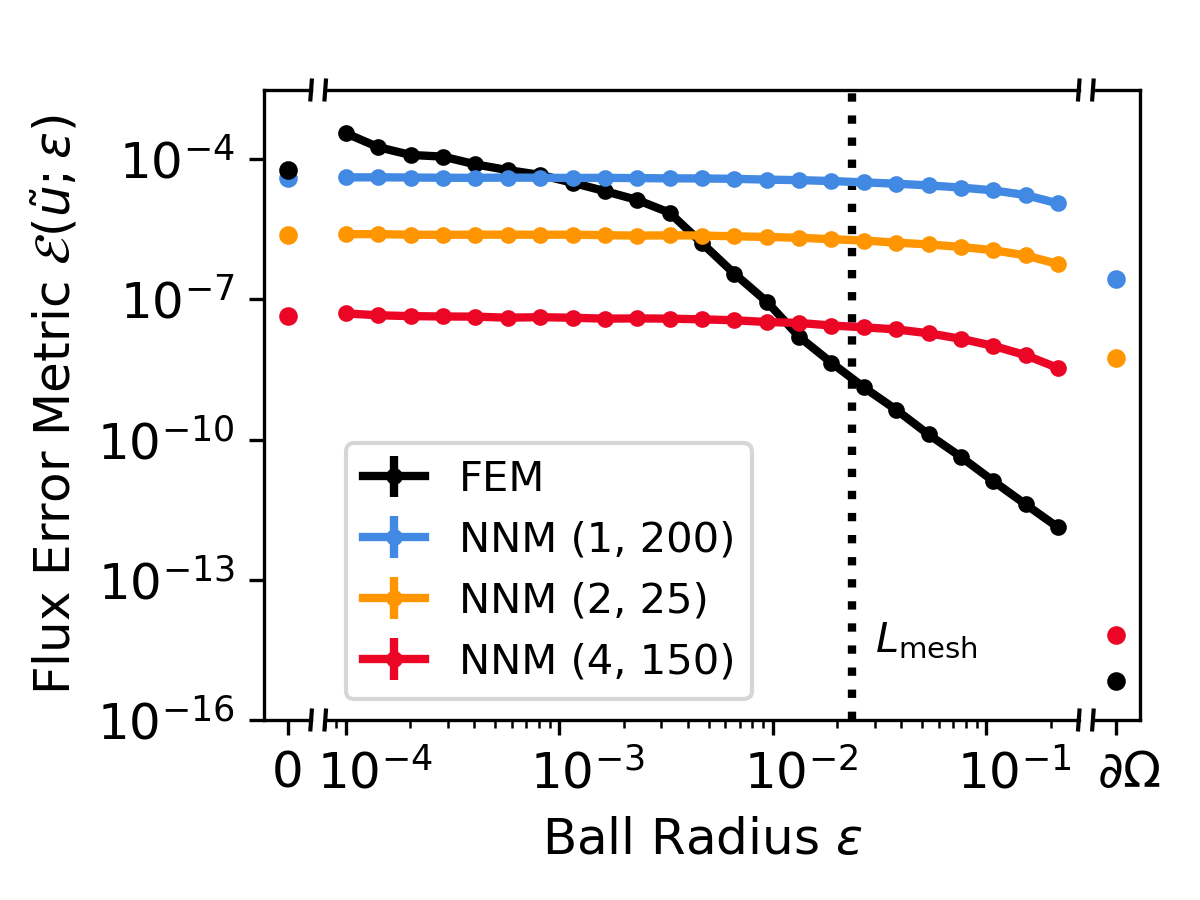}
\caption{The flux error metric $\mathcal{E}(\tilde{u};\epsilon)$ plotted as a function of the ball radius $\epsilon$ for three NNM solutions as well as the reference FEM solution. The legend entries for the NNM solutions indicate the architecture $(d,w)$ for each case. The leftmost points show $\mathcal{E}(\tilde{u};0)$, and the rightmost show $\mathcal{E}(\tilde{u};\partial\Omega)$. The dotted vertical line labeled $L_\mathrm{mesh}$ indicates the mean length scale of the FEM mesh.
 \label{fig:flux-compare}}
\end{figure}

Figure~\ref{fig:flux-compare} shows $\mathcal{E}(\tilde{u}; \epsilon)$ computed for a sample of NNM solutions (coloured lines) as well as for the reference FEM solution (black line).
The architectures, losses, and relative errors of the three networks shown in Fig.~\ref{fig:flux-compare} are listed in Table.~\ref{tbl:part-nnms}.
The shape of $\mathcal{E}(\tilde{u}; \epsilon)$ measured for the NNM solutions in Fig.~\ref{fig:flux-compare} is representative of what was measured on several other NNM solutions (not included).
In particular, $\mathcal{E}(\tilde{u}; \epsilon)$ was consistently observed to decrease monotonically with increasing $\epsilon$.
In Fig.~\ref{fig:flux-compare}, the network with architecture $(d,w)=(2,25)$ achieved relatively mediocre performance.
The $(1,200)$ network performed fairly poorly overall, but was still among the best performing shallow networks in the dataset.
As expected, the best of the three networks according to testing loss and the relative error metrics, $(4,150)$, also performed best in terms of conservation of flux.
Similarly, $(2,25)$ outperformed $(1,200)$.
We emphasize that the $(2,25)$ network outperforms the $(1,200)$ network in all metrics, despite having slightly \textit{smaller} capacity.
This is reflective of the disproportionately poor performance of shallow architectures noted in Secs.~\ref{sec:results-standard} and \ref{sec:results-arch}.

\def\arraystretch{1.5}
\begin{table}
\begin{tabular}{c|c|c|c|c|c}
$d$ & $w$ & Capacity & $\mathcal{L}[\tilde{u}]$ & $\delta_u[\tilde{u}]$ & $\delta_{\mathbf{E}}[\mathbf{\tilde{E}}]$ \\
\hline
1 & 200 & 801 & $3\cdot 10^{-3}$ & $16\,\%$ & $7.4\,\%$ \\
2 & 25 & 751 & $2 \cdot 10^{-4}$ & $1.7\,\%$ & $0.8\,\%$\\
4 & 150 & 68551 & $6\cdot 10^{-6}$ & $0.02\,\%$ & $0.08\,\%$ \\
\hline
\end{tabular}
\caption{Summary of the NNM solutions selected for the conservation of flux and particle simulations tests. Columns shown the depth, width, capacity, testing loss, and relative error of the electric potential and electric field, for each network. \label{tbl:part-nnms}}
\end{table}

The behaviour of $\mathcal{E}(\tilde{u}; \epsilon)$ for the FEM solution differs from that of the NNM solutions in some important ways.
Whereas, for all three NNM solutions, $\mathcal{E}(\tilde{u}; \epsilon)$ is roughly constant below $\epsilon \approx 10^{-1}$, for the FEM solution $\mathcal{E}(\tilde{u}; \epsilon)$ continues to increase with decreasing $\epsilon$ until at least $\epsilon \approx 10^{-4}$.
As a result, although the FEM solution achieves better $\mathcal{E}(\tilde{u}; \epsilon)$ than all NNM solutions at long length scales, the converse is true at sufficiently small length scales.
The best NNM solution in Fig.~\ref{fig:flux-compare}, $(4,150)$, exhibits comparable conservation of flux to the FEM solution at length scales near the mean FEM mesh size $L_\mathrm{mesh}=\sqrt{|\Omega|/N}$, where $N$ is the number of mesh elements.
At length scales below $L_\mathrm{mesh}$, the $(4,150)$ network conserves flux more accurately than the FEM solution.
Even the worst of the three NNM solutions shown in Fig.~\ref{fig:flux-compare} performs comparably to the FEM solution in conservation of flux at length scales below $\epsilon \approx 10^{-3}$.
The relative stability of the NNM at small length scales may be attributable to its mesh-free nature, and is an appealing feature for subsequent use in particle simulations.
Finally, we recall (see Sec.~\ref{sec:method-fem}) that the FEM solution was computed in double precision, and suggest that the single precision used for the NNM solutions may be a limiting factor to their performance at large length scales.

For small choices of $\epsilon$, $\mathcal{E}(\tilde{u}; \epsilon)$ converges to a measure of the strong form of conservation of flux.
By the divergence theorem, for a continuously differentiable field $\mathbf{\tilde{E}}$, the flux error metric $\mathcal{E}(\tilde{u}; \epsilon)$ can be rewritten as
\begin{align}
\mathcal{E}(\tilde{u}; \epsilon) &= \frac{1}{|\Omega^\epsilon|}\int_{\Omega^\epsilon} \left[ \frac{1}{|B_{\epsilon}|}\int_{B(\mathbf{x};\epsilon)} \nabla \cdot \mathbf{\tilde{E}} \,dA' \right]^2 dA, \label{eqn:div-based-flux-loss} \\
 &=  \underset{\Omega^\epsilon}{\mathrm{mean}} \left[ \left( \underset{B(\mathbf{x};\epsilon)}{\mathrm{mean}} (\nabla \cdot \mathbf{\tilde{E}}) \right)^2 \right].
\end{align}
In the remainder of this section, angle brackets $\langle \cdot \rangle_S$ will be used to denote means over any set $S$.
From Eqn.~\ref{eqn:div-based-flux-loss}, it is easy to deduce the limit of $\mathcal{E}(\tilde{u}; \epsilon)$ as $\epsilon\to 0$, which will be denoted $\mathcal{E}(\tilde{u}; 0)$.
Since $\Omega^\epsilon \to \Omega$ and the mean over $B(\mathbf{x};\epsilon)$ approaches the identity operator, it follows that
\begin{align}
\mathcal{E}(\tilde{u}; 0) &= \left\langle \left( \nabla \cdot \mathbf{\tilde{E}} \right)^2 \right\rangle_\Omega = \left\langle \left( \nabla^2 \tilde{u} \right)^2 \right\rangle_\Omega. \label{eqn:left-limit}
\end{align}
The leftmost points in Fig.~\ref{fig:flux-compare} illustrate $\mathcal{E}(\tilde{u}; 0)$ for each of the solutions.
For the NNM solutions, $\mathcal{E}(\tilde{u}; \epsilon)$ converges to $\mathcal{E}(\tilde{u}; 0)$ as $\epsilon \to 0$, as expected.
This is not the case for the FEM solution, for which $\mathcal{E}(\tilde{u}; \epsilon)$ exceeds $\mathcal{E}(\tilde{u}; 0)$ for small $\epsilon$.
However, this is not a contradiction, as Eqn.~\ref{eqn:left-limit} was derived by assuming continuous differentiability.

Equation~\ref{eqn:left-limit} is precisely the mean of the square deviation of $\tilde{u}$ from the strong form of conservation of flux.
For NNM solutions, $\mathcal{E}(\tilde{u}; 0)$ is equal to the first term of the loss functional (Eqn.~\ref{eqn:loss}) divided by $|\Omega|$, and is therefore bounded above by the loss.
Given that $\mathcal{E}(\tilde{u}; \epsilon)$ was observed to decrease monotonically with $\epsilon$, this suggests that, as for the relative errors and symmetry errors, the loss provides a useful estimator of the error in conservation of flux over any length scale.

However, as $\epsilon$ increases, the metric $\mathcal{E}(\tilde{u}; \epsilon)$ becomes increasingly biased, because the center of the balls $B(\mathbf{x};\epsilon)$ cannot be placed within a distance $\epsilon$ of the boundaries of the domain.
At moderate values of $\epsilon$, this means that errors in flux conservation in the interior of the domain are weighted more heavily than those near the boundaries of the domain.
Eventually, when $\epsilon>0.6$, the balls are too large to fit inside the slits of the device, so that only errors inside the well contribute to $\mathcal{E}(\tilde{u}; \epsilon)$.
For this reason, the data in Fig.~\ref{fig:flux-compare} are only computed for $\epsilon$ values sufficiently below 0.6 that this bias is deemed acceptably small.
This biased behaviour of $\mathcal{E}(\tilde{u}; \epsilon)$ arises because the inner integral in Eqn.~\ref{eqn:div-based-flux-loss} is based on circle-shaped test sets.
A more meaningful metric of flux conservation over very long length scales can be obtained by replacing $B(\mathbf{x};\epsilon)$ with $\partial \Omega$ in Eqn.~\ref{eqn:div-based-flux-loss}.
This global flux error will be denoted $\mathcal{E}(\tilde{u}; \partial \Omega)$, and satisfies
\begin{align}
\mathcal{E}(\tilde{u}; \partial \Omega) = \left[ \frac{|\partial\Omega|}{|\Omega|} \left\langle \tilde{E}_{\hat{n}} \right\rangle_{\partial\Omega} \right]^2 = \left[ \left\langle \nabla^2 \tilde{u} \right\rangle_{\Omega} \right]^2. \label{eqn:right-limit}
\end{align}
Thus, $\mathcal{E}(\tilde{u}; \partial \Omega)$ is directly connected to $\left\langle \tilde{E}_{\hat{n}} \right\rangle_{\partial\Omega}$, the net flux through $\partial \Omega$, which is zero for the true solution.
Note that the second equality in Eqn.~\ref{eqn:right-limit} follows from the divergence theorem, so it applies to the NNM solutions but not the FEM solution.
Together with the second equality of Eqn.~\ref{eqn:left-limit}, this means
\begin{align}
\mathcal{E}(\tilde{u}; 0) - \mathcal{E}(\tilde{u}; \partial \Omega) = \left\langle \left( \nabla^2 \tilde{u} \right)^2 \right\rangle_\Omega - \left[ \left\langle \nabla^2 \tilde{u} \right\rangle_{\Omega} \right]^2,
\end{align}
which is the variance of $\nabla^2 u$ over $\Omega$.
This is always non-negative, so it follows that
\begin{align}
\mathcal{E}(\tilde{u}; 0) \geq \mathcal{E}(\tilde{u}; \partial \Omega),
\end{align}
for any $\tilde{u}$ satisfying the second inequalities in both Eqns.~\ref{eqn:left-limit} and \ref{eqn:right-limit}.

\begin{figure}
\includegraphics[width=\columnwidth]{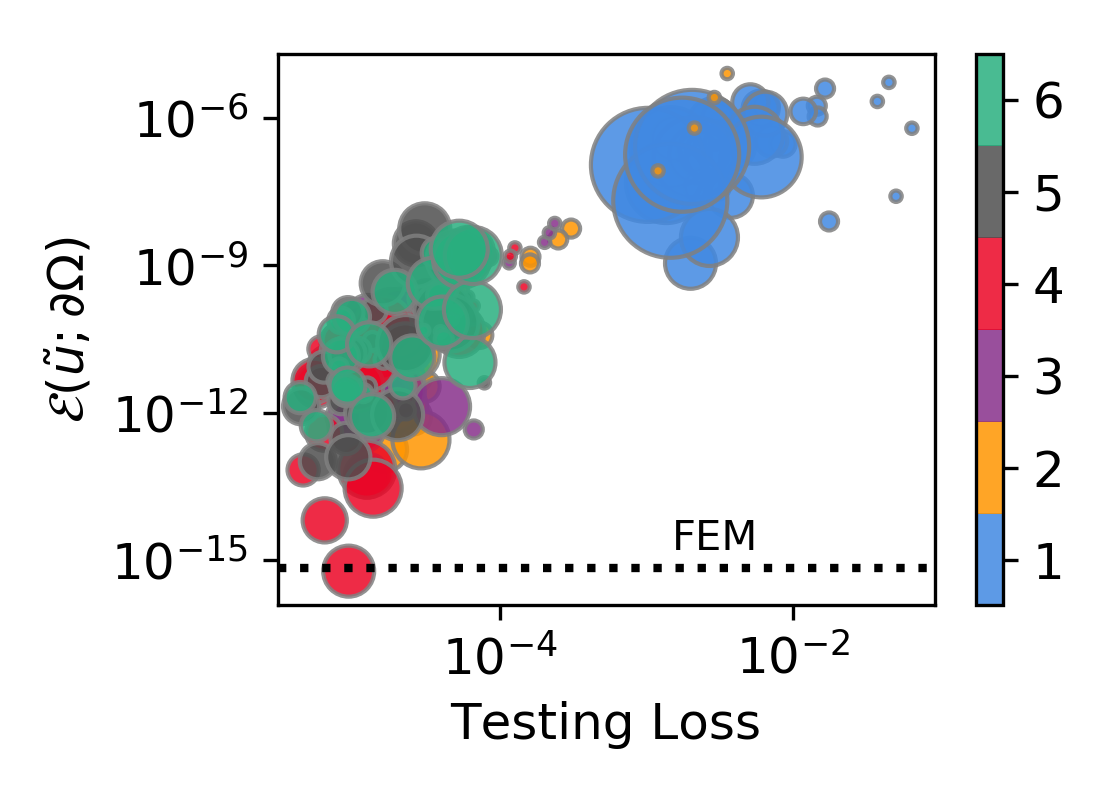}
\caption{Error in global flux conservation for all NNM solutions as a function of each network's testing loss. Marker colour indicates the depth of the network, and marker area indicates its width. The dotted line indicates the corresponding error in the FEM solution.
 \label{fig:globalflux}}
\end{figure}

The rightmost points in Fig.~\ref{fig:flux-compare} illustrate $\mathcal{E}(\tilde{u}; \partial \Omega)$ for each of the four solutions.
Figure~\ref{fig:globalflux} shows $\mathcal{E}(\tilde{u}; \partial \Omega)$ for all NNM solutions versus each network's testing loss; the dotted line indicates the value for the FEM solution.
It is immediately evident that $\mathcal{E}(\tilde{u}; \partial \Omega)$ relates to testing loss in a similar way as do the relative error metrics (Fig.~\ref{fig:err_v_loss}).
As was the case for the other metrics, $\mathcal{E}(\tilde{u}; \partial \Omega)$ decreases with decreasing testing loss, suggesting that testing loss is a useful estimator of global flux error.
Indeed, this is inevitable in the limit of small loss, since $\mathcal{E}(\tilde{u}; \partial \Omega)$ is bounded above by $\mathcal{E}(\tilde{u}; 0)$, which is in turn bounded above by the loss.
It also appears that the data in Fig.~\ref{fig:globalflux} are divided into the same two clusters as the data in Fig.~\ref{fig:err_v_loss}, with the shallow architectures performing worse than nearly all deep architectures.

Somewhat surprisingly, the best of the NNM solutions appear to conserve flux globally to nearly the same degree as the reference FEM solution, despite being computed in single (rather than double) precision.
Indeed, one network with architecture $(4,200)$ appears to slightly outperform the FEM solution in this respect.
However, it is important to note that $\mathcal{E}(\tilde{u}; \epsilon)$ for this $(4,200)$ network (not shown) exhibits essentially the same behaviour as that of the $(4,150)$ network analysed in Fig.~\ref{fig:flux-compare}.
In other words, although that particular network performs very well at global flux conservation, FEM does a significantly better job at conserving flux over intermediate length scales.
This suggests that, for the NNM solutions, the error in conservation of flux is heterogeneously distributed throughout the domain, which is consistent with the previous observation that error in the NNM solutions is significantly larger near the re-entrant corners.

In summary, the metric $\mathcal{E}(\tilde{u}; \epsilon)$ provides a mesh-agnostic measure of how well an NNM solution conserves flux over a length scale $\epsilon$.
As $\epsilon \to 0$, the limit satisfies Eqn.~\ref{eqn:left-limit}, and is bounded above by the loss.
Empirically, $\mathcal{E}(\tilde{u}; \epsilon)$ is observed to decrease monotonically with $\epsilon$, so that the loss provides a useful estimator of the error in flux conservation over intermediate length scales, too.
Alas, when $\epsilon$ is large relative to other length scales in the domain, $\mathcal{E}(\tilde{u}; \epsilon)$ is a biased metric, as it places less weight on flux lost near the boundaries of the domain.
However, a related measure of global conservation of flux over the entire domain is given by Eqn.~\ref{eqn:right-limit}, which is not biased.
This measure, too, is bounded above by the loss.
Altogether, the NNM seems capable of reliably producing solutions that conserve flux to an acceptable level of accuracy without the need to explicitly enforce this physical invariant during training.
In particular, some of the NNM solutions conserve flux globally roughly as well as the FEM solution.
Furthermore, even relatively mediocre NNM solutions conserve flux better than the FEM solution over sufficiently small length scales.

\subsection{Application to particle simulations \label{sec:results-particles}}

Section~\ref{sec:results-standard} looked directly at error between NNM and FEM, and Sec.~\ref{sec:results-special} looked at error metrics motivated by physical invariants.
Both suggested that the testing loss provides a reliable estimator of the true performance of the network solutions, and that (with appropriate network architectures) the NNM consistently finds solutions with seemingly small error values.
However, the question of what error values are acceptable is subjective, and often depends on the intended application of the numerical solutions.
For this reason, this section will consider the performance of the NNM solutions when used as the driving force fields in particle simulations  of Brownian motion in the slit-well device (implemented in C).
The simulation scenario is quite similar to those investigated by \citet{Cheng2008} and \citet{Wang2019}.

Simulations of $N=100,000$ particles in the slit-well domain were initialized with all the particles located in the middle of the same well.
The particle positions $\textbf{x}_i$ evolved according to the discretized Brownian equation,
\begin{align}
\frac{\Delta \textbf{x}_i}{\Delta t} &= \sqrt{\frac{2 D}{\Delta t}} \textbf{R}(t) + \frac{q}{\gamma} \mathbf{\tilde{E}}, \label{eqn:bd}
\end{align}
where the timestep was set to $\Delta t = 10^{-4}$, the diffusion coefficient to $D=1$, and the friction coefficient to $\gamma=1$.
The particle charge, $q$, was varied from 1 to 10.
The term $\textbf{R}(t)$ is a random driving force, representing thermal motion of an implicit solvent, and was sampled via the Box-Muller transform from an independent standard Gaussian distribution for each particle at each timestep.

The driving electric field, $\mathbf{\tilde{E}}$, was obtained from either the reference FEM solution or from one of the NNM solutions.
The electric fields were discretized onto a uniform square mesh overlain on $[-L_x,L_x]\times[-L_y,L_y]$, the smallest bounding box containing $\Omega$ (see Sec.~\ref{sec:method-nnm}).
The sidelengths of the mesh elements were set to 0.01.
The field experienced by a particle at a given position was approximated by nearest-neighbour interpolation to the mesh.
We leave more sophisticated coupling between the particle simulations and the electric fields to future work.

Particles experienced periodic boundary conditions across the left and right sides of the periodic sub-unit illustrated in Fig.~\ref{fig:slit-well-pde}, and the boundaries that were insulating in the electric field problem were treated as reflective in the particle simulations.
The number of times each particle crossed the domain was tracked, so as to measure its absolute displacement from the original position.
After $t_{\mathrm{max}} = 10^6$ timesteps, the mean  horizontal displacement of the particles from the initial position, $\langle x \rangle$, was divided by $t_{\mathrm{max}}$ to obtain an estimate $\langle v_x\rangle$ of the average particle velocity.
This average velocity was then divided by particle charge to estimate the effective particle mobility, $\mu = \langle v_x\rangle / q$.
The statistical error on this mobility measurement was estimated as $s = (\sigma_{v_x}/q) / \sqrt{N}$, where $\sigma_{v_x}$ is the standard deviation of the particle velocities.

\begin{figure}
\includegraphics[width=\columnwidth]{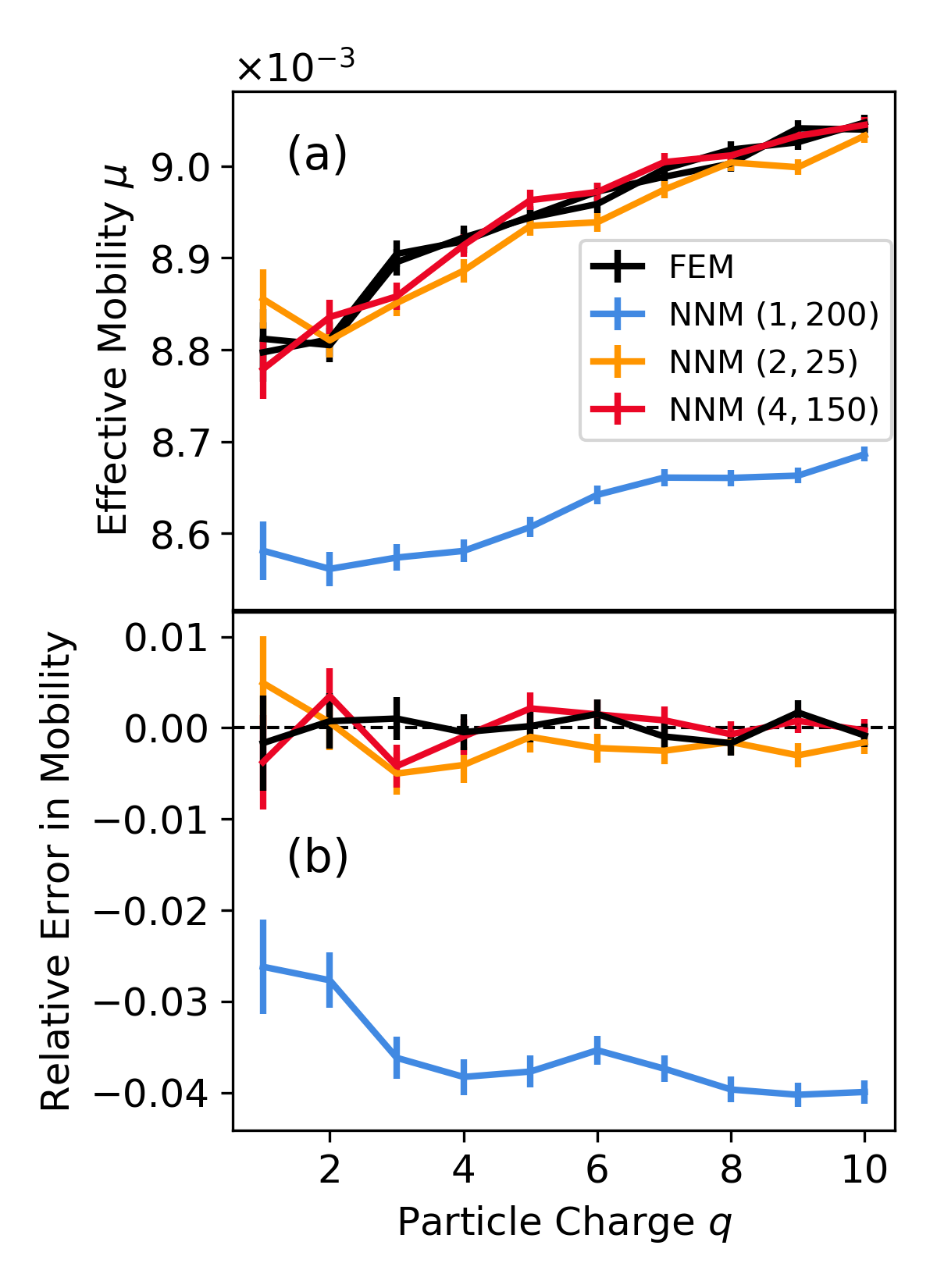}
\caption{(a) Lines show the mobility measurements $\mu$ made using four different electric field solutions. The two black lines correspond to separate simulations made using the same reference FEM field. The error bars indicate the estimated statistical error of mobility, $s$. (b) The coloured lines show the relative errors between the NNM-based measurements and the first set of FEM-based results. The black line shows the relative errors between the two sets of FEM-based measurements. The error bars are obtained from the data in (a) via standard rules for propagation of uncertainty.
 \label{fig:bd-results}}
\end{figure}

These mobility measurements are shown in Fig.~\ref{fig:bd-results}(a) for simulations conducted with the same four electric fields investigated in Sec.~\ref{sec:results-special-flux}: that of the reference FEM solution, and that of the three NNM solutions summarized in Table~\ref{tbl:part-nnms}.
The simulations using the FEM field were conducted twice with different random seeds, shown as the two black lines in Fig.~\ref{fig:bd-results}(a).
The difference between these two sets of measurements provides a means of distinguishing the errors introduced by the electric fields from simple statistical fluctuations on the mobility measurements.
In Fig.~\ref{fig:bd-results}(a), the measurements of $\mu$ made using the networks of architectures $(2,25)$ and $(4,150)$ appear fairly similar to those made using the FEM field.
Conversely, the measurements using the $(1,200)$ architecture are quite easily distinguished from the FEM data.
All simulations recovered effective mobilities that varied with $q$, induced in the otherwise free-draining particles by their interactions with the slit-well geometry.

The relative error between two mobility measurements $\mu_1$ and $\mu_2$ was quantified as
\begin{align}
\frac{\mu_1 - \mu_2}{\mu_2}.
\end{align}
The coloured lines in Fig.~\ref{fig:bd-results}(b) shows the relative errors of the NNM-based mobility measurements in Fig.~\ref{fig:bd-results}(a) versus the first set of FEM-based measurements.
The black line correspond to the relative errors between the two sets of FEM-based measurements.
Error bars were estimated via standard rules for propagation of error.

Unsurprisingly, the errors of the $(1,200)$ architecture are significantly larger than those of the other two architectures, and show a clear bias towards underestimating the mobility.
Nonetheless, even this crude solution produces errors smaller in magnitude than 5\% of the actual mobility.
This suggests that the current particle simulations are relatively insensitive to moderate inaccuracies in the driving electric field.

The relative errors of both the $(2,25)$ and $(4,150)$ architectures are comparable to the relative errors between the two sets of FEM-based measurements, and lie below 1\% for all values of $q$.
However, the relative errors for the $(2,25)$ architecture are negative for all $q$ above 2, whereas the relative errors of the $(4,150)$ architecture are roughly evenly distributed about 0.
This suggests that the $(2,25)$ architecture introduces a small but detectable systematic bias into the mobility measurements.
Conversely, the errors of the better $(4,150)$ architecture are comparable to statistical fluctuations, despite the relatively large number of simulated particles, $N=100,000$.
These results confirm that the best of the NNM solutions presented in the current work are sufficiently accurate for use in particle simulation applications.
Moreover, the relative performance of the three architectures is consistent with their values of $\mathcal{L}[\tilde{u}]$, $\delta_u[\tilde{u}]$, and $\delta_{\mathbf{E}}[\mathbf{\tilde{E}}]$ (Tbl.~\ref{tbl:part-nnms}).

In Fig.~\ref{fig:bd-results}, the network with architecture $(2,25)$ significantly outperforms that with architecture $(1,200)$, despite having slightly smaller capacity, re-emphasizing the advantages of deep architectures over shallow ones.
Conversely, the much larger $(4,150)$ architecture only achieves moderate improvements over the $(2,25)$ architecture, reflecting the diminishing returns associated with increasing network capacity.
These subtle impacts of architecture are investigated more closely in Sec.~\ref{sec:results-arch}.

\subsection{Effect of network architecture \label{sec:results-arch}}

The previous sections have demonstrated that the testing loss is a useful estimator of several independent error metrics.
Specifically, the loss functional appears to reliably estimate the error relative to the reference FEM solution; the deviation from symmetry; the deviation from conservation of flux; and the error introduced into subsequent mobility measurements.
Thus, the loss is a useful single metric of performance via which to compare different NNM architectures.

\begin{figure}
\includegraphics[width=\columnwidth]{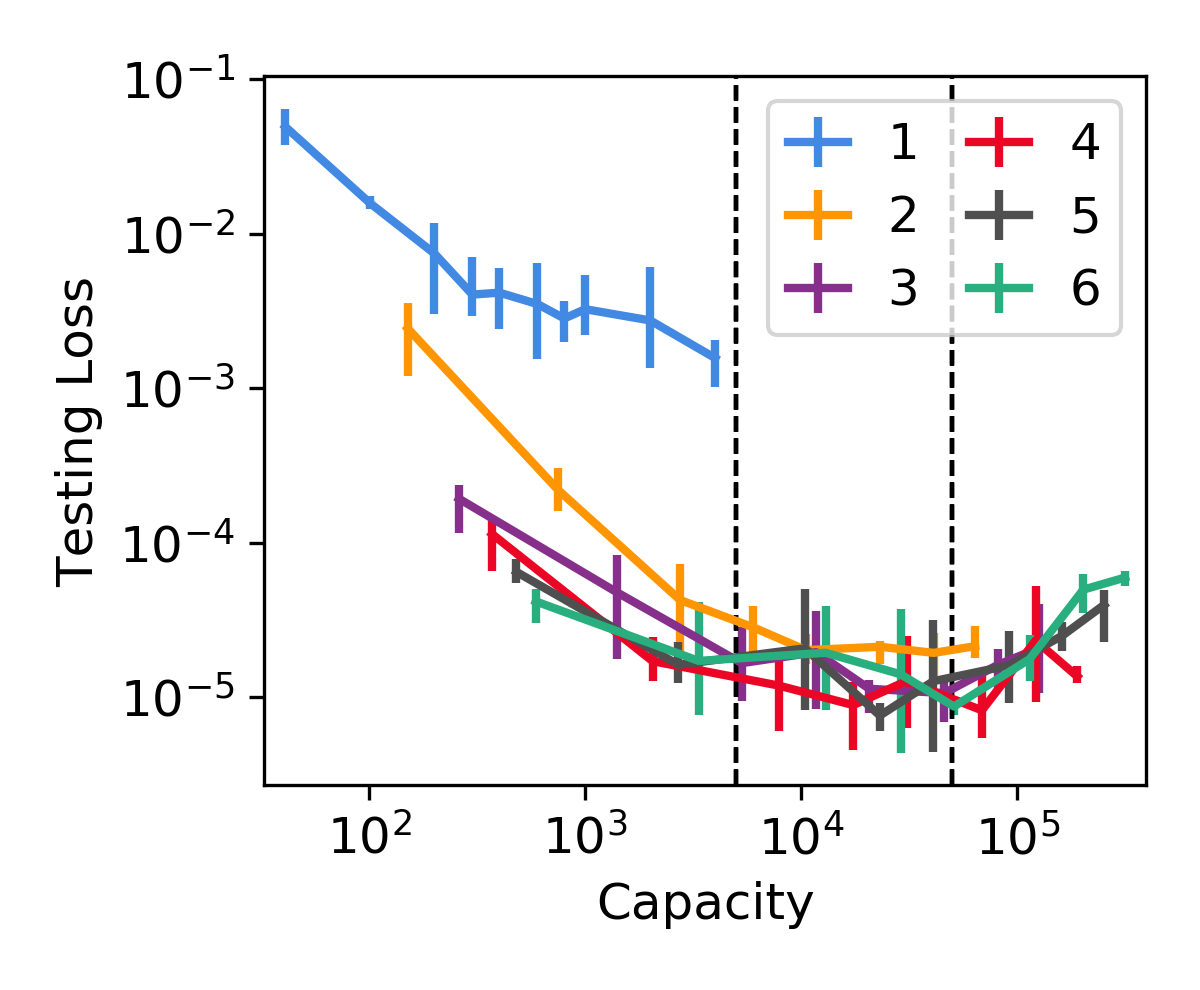}
\caption{Testing loss versus network capacity, coloured by network depths. The error bars show maxima and minima over four random seeds, and the lines indicate mean performance. The dotted lines at capacities of $5\cdot 10^3$ and $5 \cdot 10^4$ roughly delineate the three regimes discussed in the text.
 \label{fig:loss_v_capacity}}
\end{figure}

In Fig.~\ref{fig:loss_v_capacity}, the testing loss is plotted against the total network capacity.
Here, network capacity is measured as the total number of parameters in the network, given in terms of the width $w$ and depth $d$ by
\begin{align}
(2+1)w + (d-1)(w+1)w + (w+1),
\end{align}
since the networks have two inputs and one output.
The coloured lines in Fig.~\ref{fig:loss_v_capacity} correspond to different network depths, so that the various capacities within each line identify the network widths.
The error bars show maxima and minima over all random seeds, whereas the lines indicate mean performance.

The data in Fig.~\ref{fig:loss_v_capacity} show that, for network capacities below $5 \cdot 10^3$, increasing capacity improves testing loss for any choice of depth.
This suggests that, for those networks, insufficient capacity is a primary bottleneck towards representing more accurate approximations of the true solution.
In particular, for the networks with two hidden layers, increasing the capacity improves the loss by nearly two orders of magnitude.
Furthermore, in this low-capacity regime, increasing depth improves performance for a given capacity.
In other words, when insufficient network capacity is the primary barrier to improved performance, deeper networks make more efficient use of that limited resource.
Indeed, this is consistent with the effects of architecture observed in Figs.~\ref{fig:err_v_loss}, \ref{fig:flux-compare}, \ref{fig:globalflux}, and \ref{fig:bd-results}.
Specifically, shallow networks perform particularly poorly in all metrics throughout the present work, even compared to networks with comparable capacity and as few as two hidden layers.

For deep networks with moderately large capacities ($5\cdot 10^3$ to $5 \cdot 10^4$), testing loss is essentially independent of network architecture (i.e., independent of both depth and capacity/width).
This suggests that insufficient network capacity is no longer a primary bottleneck to improving solution accuracy.
The investigation by \citet{Magill2018neurips} suggested that the internal representations learned by networks in the NNM become essentially independent of width above some critical size, so it is not surprising that loss similarly becomes independent of width.
However, it is noteworthy that this limiting loss value is also independent of network depth (among those with two or more hidden layers).

For networks with capacities of $5 \cdot 10^4$ or above, testing loss begins to increase with further increases in capacity.
Figure~\ref{fig:err_v_loss} illustrates that these same networks sometimes exhibit relative errors nearly as high as some shallow networks, despite having two orders of magnitude more capacity.
Their poor performance can be understood in terms of the difficulties commonly encountered in training very deep, wide neural networks.
For instance, \citet{Berg2018} noted similar loss in performance when training networks with five or more hidden layers, and attributed this to vanishing gradients.
Refinements in the network architectures and training algorithms can be expected to alleviate this phenomenon.

Note that the behaviour of these networks with very large capacities cannot be described in terms of overfitting, another problem commonly encountered by networks with excessively large capacities.
Overfitting is typically defined as a significant gap between the training and testing losses of networks.
In the NNM, however, the testing and training sets are drawn from identical distributions.
In the implementation used here, in particular, the training set is redrawn regularly throughout training, so that it is fundamentally impossible for the network to be overfitting to a specific set of training samples.

\begin{figure}
\includegraphics[width=\columnwidth]{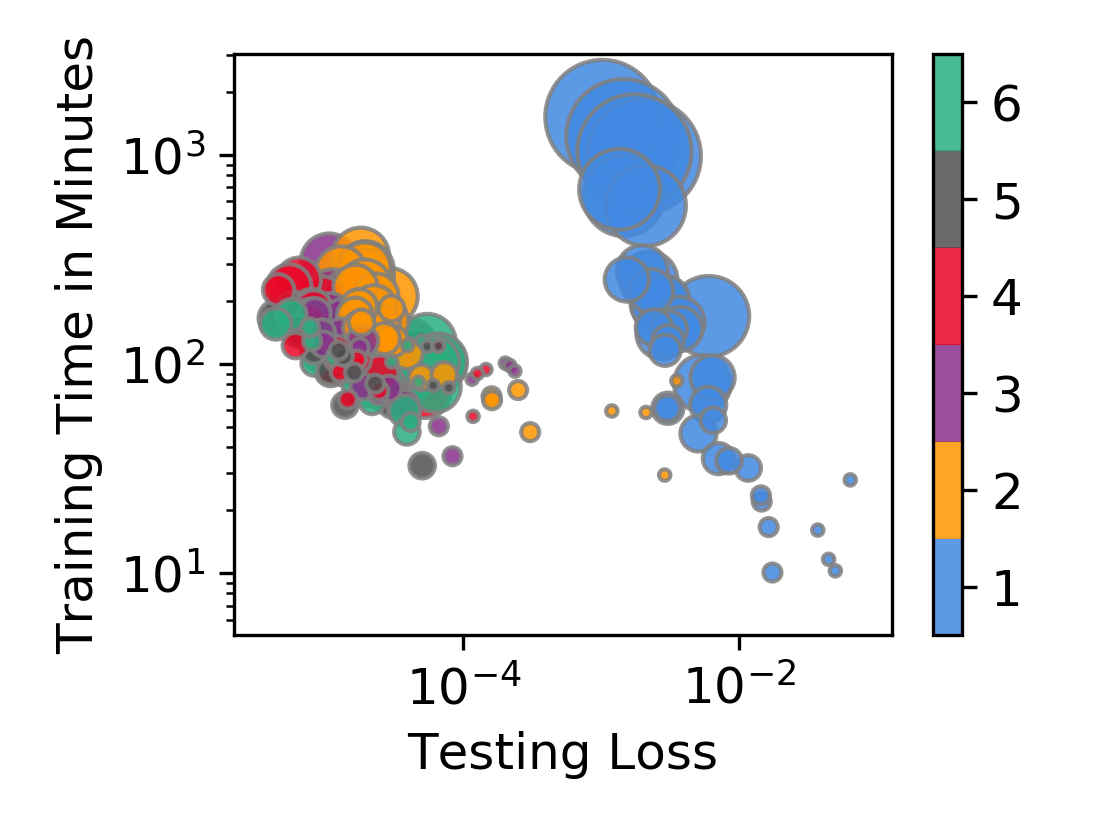}
\caption{Total training time and final testing loss of the NNM solutions. Marker colour indicates the depth of the network, and marker area indicates its width.
 \label{fig:time_v_loss}}
\end{figure}

Finally, Fig.~\ref{fig:time_v_loss} shows the total training time of the NNM solutions against testing loss.
The same two populations identified in Figs.~\ref{fig:err_v_loss} and \ref{fig:globalflux} are evident again in Fig.~\ref{fig:time_v_loss}.
The cluster on the right contains all the shallow networks as well as the narrowest of the deep ones.
The cluster on the left consists of those networks that attained better than 1\% error relative to FEM (Fig.~\ref{fig:err_v_loss}).
Within each cluster, testing loss and training time are loosely correlated.
For all networks, training time was on the order of hours.
However, it is important to note that the implementation in the present work was not concerned with optimizing the computational efficiency of the NNM, but rather with ensuring that the training process was thoroughly converged (Sec.~\ref{sec:method-nnm}).

Once again, the networks in the right cluster perform disproportionately poorly, even though many of them have capacities comparable to some of those in the left cluter (Fig.~\ref{fig:loss_v_capacity}).
Thus, not only do the networks in the left cluster achieve better accuracies (as measured by testing loss or any of the various error metrics in this paper), but they also finish training far more rapidly.
Further, this conclusion is true even between networks of equal capacity.
These observations demonstrate many benefits of using deeper architectures in the NNM, and several disadvantages of using shallow architectures.

\section{Conclusions}

This work investigated the performance of the neural network method (NNM) when used to solve the electric potential and field in the slit-well device.
This problem features a non-convex geometry, which makes it particularly challenging to solve with the NNM.
Performance was quantified in multiple metrics, and compared against a reference FEM solution.

The best network architectures studied here reliably achieved relative errors below 0.1\% in both the potential and the field.
NNM solutions also recovered spatial symmetries of the true solution to roughly the same extent that they approximated the true solution.
Regarding conservation of flux, the NNM solutions performed comparably to the reference FEM solution.
Finally, particle simulations conducted using the NNM electric fields yielded mobility measurements consistent with those based on the FEM electric field.
In each of these metrics, the testing loss was found to provide a useful estimator of the networks' true performance.
That is, networks with smaller losses were found to be closer to the true solution; to more closely approximate the target symmetries; to conserve flux more accurately; and to produce better particle simulations.

These empirical investigations uncovered several valuable insights for practical use of the NNM.
Accurate solutions to physical problems can be obtained even without explicitly enforcing known physical invariants of the true problem.
The importance of architecture was re-emphasized: deep architectures consistently outperformed shallow ones, converging to better solutions in less time and using fewer degrees of freedom.
Finally, the testing loss may provide a practical means of gauging a solution's accuracy, even when the ground truth is unknown and convergence is not theoretically guaranteed.

In summary, this work demonstrates that the NNM can successfully solve a problem that is ill-conditioned due to the non-convexity of its domain.
The NNM solutions were found to be particularly appropriate for use in subsequent particle simulations.
This suggests that it could be a useful tool for the study of micro- and nanofluidic devices (MNFDs) and other biophysical systems.
Moreover, differential equations in domains with complicated geometries arise throughout physics and other fields.
These results support the feasibility of using the NNM to solve this fundamental and ubiquitous class of problems.

\appendix

\section{Additional plots of the electric field solution}

\begin{figure}
\vspace{2em}
\includegraphics[width=0.9\columnwidth]{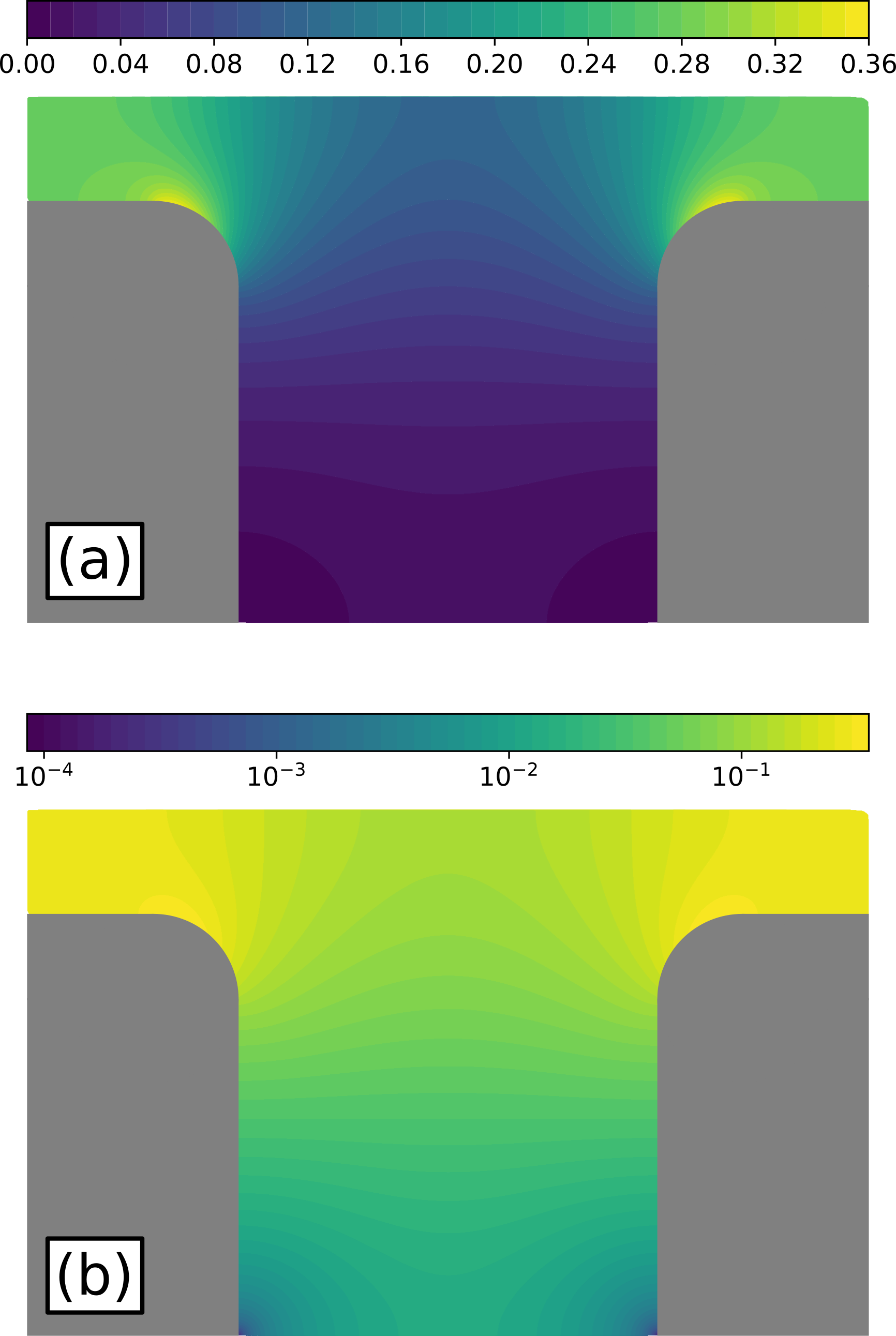}
\caption{Electric field intensity of the FEM solution, shown on (a) linear and (b) logarithmic colour scales.
\label{fig:slit-well-absE}
}
\end{figure}

Figure~\ref{fig:slit-well-absE} shows the FEM electric field intensity throughout the domain, in both linear and logarithmic colour scales.
In particular, Fig.~\ref{fig:slit-well-absE} illustrates that the peak field intensity occurs near the re-entrant corners, with a magnitude of about 0.36.
In the bottom corners of the well, the field intensity is over four orders of magnitude weaker.
These features contribute to the difficulty of applying the NNM to the slit-well electric field problem, since the standard loss functional used during training places equal weight on all regions of $\Omega$ and $\partial \Omega$.
The regions of very intense electric field near the re-entrant corners, specifically, seem to be most difficult to resolve for the NNM, as seen in the error maps shown in Fig.~\ref{fig:slit-well-soln}.

\section{Details of flux loss calculations \label{app:fluxcalc}}

This appendix contains descriptions of how the metrics shown in Figs.~\ref{fig:flux-compare} and \ref{fig:globalflux} were computed.
For Fig.~\ref{fig:flux-compare}, the integrals in Eqn.~\ref{eqn:field-based-flux-loss} were computed by sampling 10000 uniformly spaced points on $\partial B(\mathbf{x};\epsilon)$ for each choice of the center $\mathbf{x}$.
Candidate samples for the centers were generated according to the same procedure described in Section~\ref{sec:method-nnm}, but with ten times higher sample density, and all points within a distance $\epsilon$ of $\partial \Omega$ were rejected.

The leftmost points in Fig.~\ref{fig:flux-compare} correspond to Eqn.~\ref{eqn:left-limit}.
For the NNM solutions, these were computed by Monte Carlo integration over $\Omega$ using ten times higher sampling density than in Sec.~\ref{sec:method-nnm}.
The rightmost points in Fig.~\ref{fig:flux-compare} correspond to Eqn.~\ref{eqn:right-limit}.
These were not computed using a Monte Carlo integration approach.
Because $\left\langle \tilde{E}_{\hat{n}} \right\rangle_{\partial \Omega}$ is a small number computed by summing many positive and negative terms, it is vulnerable to catastrophic cancellation.
For this reason, it was computed using a uniform mesh of points along $\partial \Omega$, sampled with 100 times higher density than in Sec.~\ref{sec:method-nnm}.
For the FEM solution, the integrals required for Eqns.~\ref{eqn:left-limit} and \ref{eqn:right-limit} were both computed in FEniCS using Gaussian quadrature via the \verb-assemble- command.

\FloatBarrier

\setlength{\bibsep}{10pt}
\bibliographystyle{unsrtnat}
\bibliography{2019-11-field-refs}

\end{document}